\documentclass{elsart}
\usepackage{ifpdf}
\usepackage{graphicx,natbib,amssymb,lineno}

\makeatother

\begin{document}
\begin{frontmatter}

\title{Two Component Baryonic-Dark Matter Structure Formation in Top-Hat Model}

\author{M. Malekjani}

\address{Department of Theoretical Physics and Astrophysics,
University of Tabriz, P.O.Box 51664, Tabriz, Iran \\
malekjani@tabrizu.ac.ir}

\author{S. Rahvar}

\address{Department of Physics, Sharif University of Technology,
P.O.Box 11365--9161, Tehran, Iran\\
rahvar@sharif.edu}

\author{D. M. Z. Jassur}

\address{Department of Theoretical Physics and Astrophysics,
University of Tabriz, P.O.Box 51664, Tabriz, Iran \\
jassur@tabrizu.ac.ir }

\begin{abstract}
In this work we extend simple top-hat model of structure formation
to the two-component system made of baryonic and dark matter. We
use Harrison-Zeldovich spectrum as the initial condition for the
structures and calculate their evolution up to the present time.
While we do not take into account some complications during the
structure formation, such as the merging of galaxies, however this
formalism can give us a qualitative picture from the formation of
structures. We show that in this model small scale structures
evolve faster than the larger ones and it predicts a down-top
scenario for the structure formation. The trend of power spectrum
in this model is compatible with the observations and results in
$\sigma_8 \sim 0.8$. This formalism provides an analytic treatment
of structure growth and can easily show the effect of the
cosmological parameters on the formation of the structures. As an
example, the effect of a parameterized dark energy model on the
growth of the structures is investigated.

\end{abstract}
\end{frontmatter}

\section{Introduction}
Observations of the Cosmic Microwave Background (CMB) by the COBE
satellite and the subsequent experiments such as WMAP indicate the
existence of temperature fluctuations of order $\sim10^{-5}$ at
the last scattering surface \cite{smo92,hin06,spe03}. This
temperature contrast on CMB may result from the primordial quantum
fluctuations at the early universe. The quantum fluctuations in
the inflationary scenario, provide a specific spectrum for the
 matter so-called Harrison-Zeldovich. However, Recent observations by
 WMAP show a small deviation from this spectrum ($n_s=0.958\pm0.016$) \cite{spe06}.
  An outstanding characteristic
of this spectrum is its scale-independent property, means that all
the perturbations have some density contrast of about $ 10^{-5}$ at
the entering time to the horizon \cite{spri05}. One of the questions
in cosmology is that of how these small perturbations at
$z_{dec}=1100$ can grow to the present non-linear structures while
we expect from the standard structure formation theory that they
should evolve to a density contrast of $\delta \sim
\delta_{dec}\times z_{dec} = 10^{-2}$ at the present time
\cite{pee80}.

Including dark matter as one of the components of the universe is
a solution to this question. Since the entering time of the
structures to the horizon depends on the size of structure, we
expect small scale structures enter earlier than the larger ones.
Meanwhile, before the decoupling epoch, pressure of the radiation
prevents formation of baryonic structures smaller than the Silk
mass \cite{sil68}, the dark matter structures continue their
growth. After decoupling, the mutual interaction of dark
matter-baryonic matter speeds up the growing rate of the structure
formation and results in non-linear structures at the present
time.

Here in this work we use two components of baryonic and dark
matter in the top-hat model for studying their growth. We take the
initial condition of sub-horizon over-dense regions at the last
scattering surface. The size of perturbations in our concern
guarantees using of Newtonian mechanics \cite{man02,rah03}. After
decoupling, the top-hat sphere starts to expand up to a maximum
radius and then turns-around to collapse. During the collapse,
once the structure satisfies the virial condition, the global
velocity turns into the dispersion velocity and thermalizes the
gas of structure and finally prevents a catastrophic collapse of
the structure. The result of thermalization is the ionization of
the baryonic gas and consequently gas starts to cool through the
radiation. Cooling makes baryonic structure to contract further
and finally the baryonic component reaches to a stable stage. We
should point out that in this scenario we ignore merging effects
during the formation of the structures and this model can be
applicable only for the isolated systems. Finally we compare
$\sigma_{8}$ from this simple theoretical model with that from
weak-lensing.

The organization of paper is as follows: In Section \ref{tophat}
we introduce Standard Collapse Model (SCM), extend it for the two
component fluid and obtain dynamics and power spectrum of the
structures. Section \ref{variable} discusses about the effect of
variable dark energy on the formation of structure in top-hat
model. In Section \ref{cooling} we study the cooling effect on the
evolution of baryonic matter after the thermalization and estimate
corresponding redshift for the star formation. We conclude in
section \ref{conc}.
\section{Spherical Top-Hat Model: Structure Formation}
\label{tophat} In this section we review the standard spherical
collapse model. Here we take a spatial uniform distribution of the
matter inside the structure, so-called top-hat distribution which
is slightly denser than the uniform background density. One of the
advantages of this model is that it has an analytical solution for
the dynamics of the structure. In this section first we introduce
the standard top-hat model and then extend it, introducing the two
component fluid in the structure: (i) a non-dissipative dark
matter and (ii) a dissipative baryonic matter.
\subsection{standard top-hat model}
A simple approach for studying the structure formation in the universe
 is the spherical collapse model. We take a spherical region embedded in
the uniform background which has a tiny density deviation from
that of the background. The scale of this region is much smaller
than horizon and the velocities are non-relativistic. These two
conditions guarantee the application of the Newtonian gravity for
studying the growth of the structures \cite{man02,rah03}.

For a spherical region with the radius R(t) and uniformly
distributed mass of $M$, containing non-relativistic matter, the
density contrast is given by:
\begin{equation}
1+\delta(t)=\frac{\rho(t)}{\rho_{b}(t)}=\frac{3M}{4\pi
R^3(t)}\frac{1}{\rho_{b}(t)}, \label{dcontrast}
\end{equation}
where $\rho_b(t)$ is the homogenous background density of the
universe. The energy and momentum equations for a non-dissipative
spherical matter is given by:
\begin{eqnarray}
\label{energy} \label{energy}
E &=& \frac12 \dot{R}(t)^2 - \frac{GM}{R(t)},\\
\ddot{R}(t)&=&-\frac{GM}{R(t)^2} \label{momentum}
\end{eqnarray}
where $E$ is energy per mass and can be calculated from the
initial condition of the structure, (i.e. $E=E_i$). The initial
radial velocity of the structure is taken by $v_i = H_iR_i +
v_{i}^{(pec)}$, where $H_i$ is the Hubble parameter of the
background and $R_i$ is the size of the structure at the initial
time, $v_{i}^{(pec)}$ is the peculiar velocity of the structure
and can be given by $v_i^{(pec)}=-H_iR_i\delta_i$ \cite{pee80}.
Using the dependence of the peculiar velocity to the density
contrast, the radial velocity of the structure at the initial time
is:
\begin{equation}
v_i = H_iR_i(1-\delta_i), \label{pecv}
\end{equation}
where $\delta_i$ depends on the size of the structure. Using
equation (\ref{pecv}), the Kinetic energy per unit mass of the
structure is $K_i = K_i^{(b)}(1-2\delta_i)$ where $ K_i^{(b)}$ is
the Kinetic energy of background at a distance $R_i$ from the
center of coordinate\footnote{Note that the effect of density
contrast in the velocity of the structure and subsequently on the
Kinetic Energy of over-dense region is missed in the text book
\cite{pad93}.}. For the initial potential energy of structure we
have $U_i = \Omega_iK_i^{(b)}(1+\delta_i)$. The total energy is
given by the sum of the kinetic and the potential energy of the
structure at the initial time as:
\begin{equation}
E = -k_i^{(b)}\Omega_i(1+\delta_i+2\delta_i\Omega_i^{-1} -
\Omega_i^{-1}). \label{e5}
\end{equation}
For the case of spatially flat universe, $E=-3k_i^{(b)}\delta_i$.
Integrating from equation (\ref{energy}) results in the equation
of motion in the parametric form:
\begin{eqnarray}
\label{tt}
R(\theta)&=&A(1-\cos\theta),\\
t(\theta)&=&T + B(\theta-\sin\theta), \label{rt}
\end{eqnarray}
where $\theta$ varies in the range of $\in[0,2\pi]$ and $T$ is a
constant. For $\theta = \pi$, the structure reaches to the maximum
radius of $r_{max} = 2A$. Substituting in equation (\ref{energy})
at the maximum radius provides $A=GM/2E$ which results in
$$ A = \frac16 \frac{R_i}{\delta_i}$$
On the other hand using
equation (\ref{momentum}) yields:
$$A^3 = GMB^2$$
Substituting equations (\ref{tt}) and (\ref{rt}) in
(\ref{dcontrast}) results the evolution of the density contrast in
terms of $\theta$ as:
\begin{equation}
\delta=\frac{9(\theta-\sin\theta)^2}{2(1-\cos\theta)^3}-1.
\label{delta}
\end{equation}
For the initial condition, considering $\delta_i\ll1$, the initial
phase is $\theta_i = 2\sqrt{\delta_i}$. Taking $T=0$, $t(\theta)$
will coincide with the cosmic time. From the initial condition, $B$
obtain as:
$$ B = \frac{3t_i}{4\delta_i^{3/2}}.$$
From equation (\ref{delta}) at $\theta = 2\pi/3$ the structure
enters to the non-linear regime (i.e. $\delta\simeq 1$).

On the other hand for $\theta = 2\pi$  we have singularity,
however before this stage, global radial velocity of the structure
is converted to the dispersion velocity and prevent it from the
catastrophic collapsing. The virial theorem provides us a radius
that the stable condition of the structure is fulfilled. In the
next part we will extend the top-hat model to the two component
fluid of the baryonic and dark matter in which they are
gravitationally coupled during the evolution of the structure.
\subsection{two component top-hat model}
Evolution of the large scale structures indicates that dark matter
is an essential element for the formation of the structures in the
universe. In the standard scenario of the structure formation,
structures composed by baryonic matter and dark matter are in
mutual gravitational interaction during their evolution. In this
section we obtain the evolution of each component in the
structure, using the top-hat model.

\begin{figure}
\centerline{\includegraphics[width=0.8\textwidth]{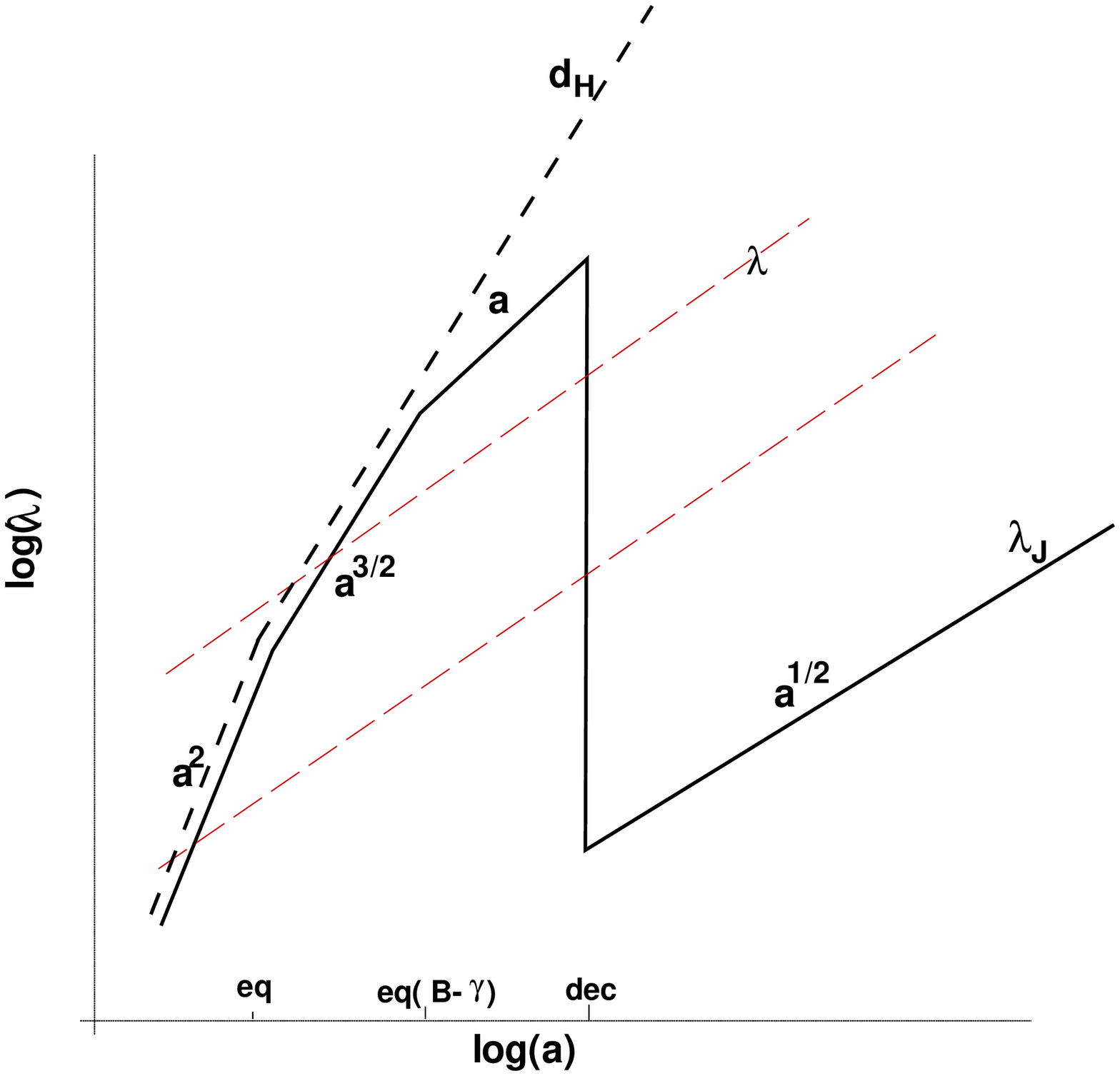}}
\caption{\small Size of horizon of universe (short dashed line),
size of structure (long dashed line) and Jeans length (solid line)
of baryonic structures in terms of scale factor in logarithmic
scale. At the last scattering surface the Jeans mass of baryonic
matter decreases and baryonic structure with sub-horizon scale can
grow after this epoch.} \label{jeanslength}
\end{figure}

Let us start with baryonic component. Baryons before the
decoupling were tightly coupled to the photons and  diffusion of
the radiation prevents them to collapse under their own gravity.
The corresponding diffusive mass of the baryonic structure with
$\lambda<l_{diff}$ is called the Silk mass with the mass of $M_S =
6.2\times 10^{12} (\Omega/\Omega_B)^{3/2} (\Omega h^2)^{-3/4}
M_{\odot}$ \cite{pad00}. However, after decoupling of the baryons
from the cosmic microwave background radiation, Jeans length
decreased rapidly and baryonic structure could grow freely (see
Fig. \ref{jeanslength}). The corresponding Jeans mass after the
decoupling is
\begin{equation}
M_{J} = \rho_b(\frac{kT}{G\rho_b m_p})^{3/2}\simeq 10^{5}M_{\odot},
\end{equation}
where all the parameters are calculated at the last scattering
surface with $T\sim 3000 K$ and $m_p$ is the mass of the proton.
The rest of the scenario is the gravitational interaction of the
baryonic matter larger than the corresponding Jeans mass with the
gravitational potentials that have already been made by the dark
matter structures.

Comparing the mass of galaxies and cluster of galaxies with the
Silk mass at the last scattering surface shows that we can set
$\delta_b<10^{-5}$ which can be ignored compare to the density
contrast of dark matter ($\delta_b \ll \delta_{d} \simeq
10^{-3}$).

We consider two spherical regions with radii of $R_{b}(t)$ and
$R_{d}(t)$, and total masses of $M_{b}$ and $M_{d}$ for the
baryonic and dark matter components of the structure. These two
spheres are coincided on each other at the initial time, but due
to the different initial conditions they will evolve with
different rates. Similar to the first part of this section, the
momentum and energy conservation equations for the dark matter is
written as
\begin{eqnarray}
\label{dark}
{\ddot R}_{d}&=&-G\frac{M_{d}+M_{b}(t)}{R_{d}^{2}}, \\
E_{d}&=&\frac{1}{2}\dot{R_{d}^{2}}-G\frac{M_{d}+M_b(t)}{R_{d}}.
\label{energy_dark}
\end{eqnarray}
We let the initial density-contrast for the baryonic matter
($\delta_b \simeq0$) which provides a zero initial peculiar
velocity for the baryonic component. Comparing to the dark matter
sphere, the baryonic sphere will expand faster. According to this
picture from the dynamics, the dark matter will interact
gravitationally only with a fraction of the baryonic matter inside
the dark matter sphere.

Similarly, the dynamics of baryonic sphere is given by:
\begin{eqnarray}
\label{baryon}
{\ddot R}_{b}&=&-G\frac{M_{d}+M_{b}}{R_{b}^{2}}, \\
E_{b}&=&\frac{1}{2}\dot{R_{b}^{2}}-G\frac{M_{d}+M_b}{R_{b}}.
\label{energy_baryon}
\end{eqnarray}
As the baryonic sphere is always larger than the dark matter
sphere, this component interacts with all the dark sphere. The
energy of the two spheres can be obtained from Eq.(\ref{e5}).
Setting zero peculiar velocity for the baryonic sphere, results in
$E_b = -k_i^{(b)}\delta_i^{(d)}$ and for the dark matter component
$E_d = -3k_i^{(b)}\delta_i^{(d)}$.

To calculate the initial condition for the density contrast of
dark matter $\delta_i^{(d)}$, we use the Harrison-Zeldovich
power-law spectrum \cite{har70,zeld70} as
\begin{equation}
P(k)=Ak^{n},
 \label{spectrum}
\end{equation}
where we adapt $n = 1$. The corresponding mass variance of this
spectrum is $\sigma^2 = \frac{A}{4\pi^2} k^4$. An important
specification of this spectrum is that at the entering time of the
structure to the horizon the density contrast has an invariant value
of $\sigma_{enter} = {2\pi A^{1/2}}/{(9t_0^2)}$. Using $A\simeq
(28.6 h^{-1} ~Mpc)^4$, the density contrast for the entering time is
$\delta_{enter} \simeq 6 \times 10^{-5}$. Our aim at this stage is
applying the Harrison-Zeldovich spectrum to have the density
contrast of the dark matter at the decoupling time. The density
contrast at the last scattering depends on the epoch that structure
enters to the horizon. We divide the evolution of density contrast
into two area of radiation and matter dominant epochs. In these two
phases, the dynamics of dark matter changes with different rates in
terms of scale factor.

\begin{figure}
\centerline{\includegraphics[width=0.8\textwidth]{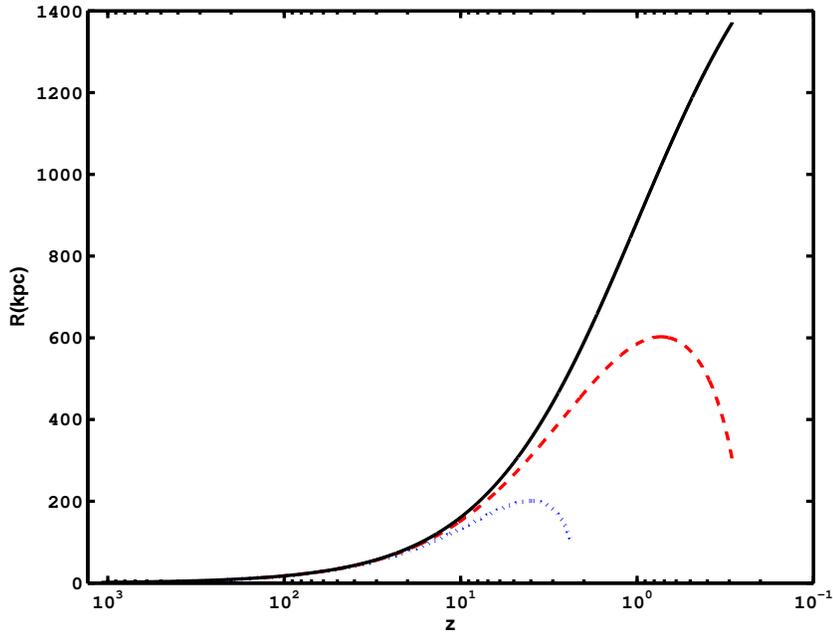}}
\caption{\small Dependence of radii of halo (dotted-line) and
baryonic (dashed-line) components for astructue with a galaxy mass
in top-hat model in terms of redshift, compared with that of
background (solid-line). Background is taken $\Lambda$CDM model
with the corresponding parameters of $\Omega_{m}^{(0)} = 0.3$,
$\Omega_{\Lambda}^{(0)} = 0.7$ and $H_0\sim 70 ~ Km/s Mpc^{-1}$.}
\label{galaxy}
\end{figure}

\begin{figure}
\centerline{\includegraphics[width=0.8\textwidth]{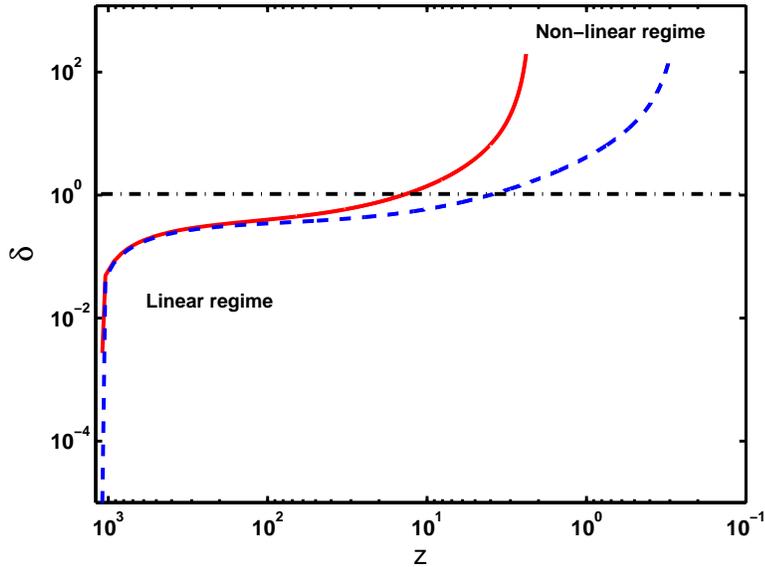}}
\caption{\small Density contrast evolution of baryonic
(dashed-line) and dark matter (solid-line) in terms of redshift.
The horizon line represents $\delta=1$, separate the liner and
non-linear regimes.}\label{dcont}
\end{figure}

Let us first consider the structures that enter the horizon at the
radiation dominant epoch. The structures having smaller than the
horizon mass at the equality time $(M<M_H(eq))$ will enter horizon
at the following redshift:
\begin{equation}
z_{enter}(M)\simeq z_{eq}(\frac{M}{M_{H}(z_{eq})})^{-\frac{1}{3}}.
\end{equation}
Letting horizon mass at the equality epoch
$M_{H}(z_{eq})\simeq5\times10^{15}(\Omega h^{2})^{-2}M_{\odot}$,
the corresponding entering redshift of a structure to the horizon
can be calculated. For instance the dark matter component of a
structure with a galaxy mass enters the horizon at
$z_{eneter}(galaxy) \sim 5.90 \times 10^4$. The structures grow
during $z_{enter}$ (enter time) to $z_{eq}$ (equality time) by a
logarithmic factor of
\begin{equation}
\delta(a_{eq})\simeq 5\ln(\frac{a_{eq}}{a_{enter}})\delta(enter),
\end{equation}
where we use $z_{eq}= 3454^{+385}_{-392}$ \cite{spe03}. Structure
with a galaxy size at the present time will grow by the factor of
$20$ during $z_{enter}$ to $z_{eq}$ period which results a density
contrast of about $\delta_{eq}(galaxy) \sim 8.50 \times 10^{-4}$
at equality time. Once the universe enters to the matter dominate
era, structure start to grow proportional to the scale factor and
at the decoupling time the density contrast of dark matter will
reach to $\delta(z_{dec}) = a_{dec}/a_{enter}\times
\delta_{enter}$.
 For a structure with the galaxy mass, the dark
matter density contrast grows up to $\delta_{dec} \simeq 2.7\times
10^{-3}$ at the decoupling time ( we use $z_{dec}= 1088^{+1}_{-2}$
\cite{spe03}). Fig.(\ref{galaxy}) shows the dynamics of the radii of
dark matter and baryonic components of a galaxy in terms of
redshift. We use $\Lambda$CDM model for the background with the
parameters of $\Omega_{m}^{(0)} = 0.3$, $\Omega_{\Lambda}^{(0)} =
0.7$ and $H_0\sim 70 ~ Km/s Mpc^{-1}$. The time dependence of
density contrast also is shown in Fig. (\ref{dcont}). The evolution
of each component is calculated up to the virilization time. We note
that the dark matter component virialize earlier than the baryonic
part.

\begin{figure}
\centerline{\includegraphics[width=0.8\textwidth]{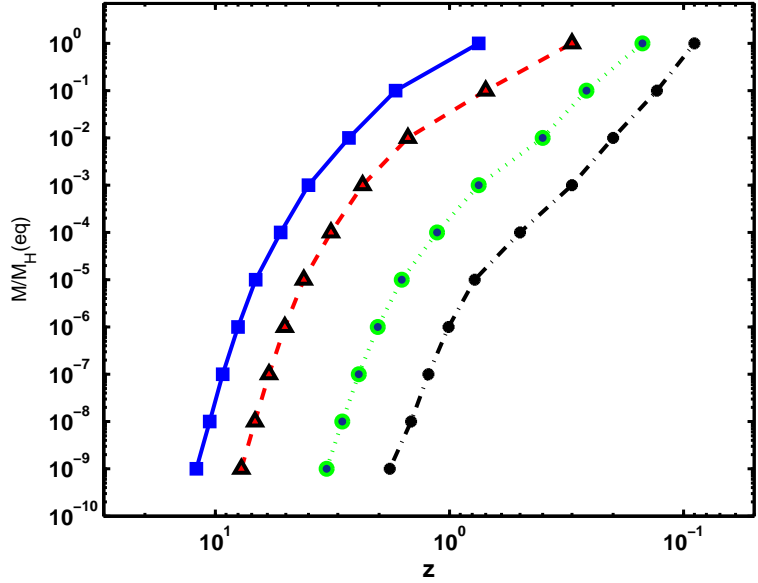}}
\caption{\small  Dependence of Mass (normalized to $M_H$)to the
characteristic redshifts of the structures. Solid line represents
redshift corresponds to the maximum radius of a structure in terms
of mass. Dashed line represents the dependence of virialization
redshift to the mass of dark matter structure. Dotted and
dashed-dotted lines represent the maximum radius and virialized
redshifts for the baryonic component of the
structure,respectively.} \label{mvv}
\end{figure}

We calculate the evolution of the gravitationally coupled two
component in the top-hat model, for the structures in the range of
$10^6$ to $10^{15}$ solar masses. These structures, depending on
their masses, will enter the horizon at different epochs, smaller
sooner and the larger later. Means that the smaller structures
grow faster than larger ones (see Table \ref{tab1} and Fig.
\ref{mvv}). This result is in agreement with the down-top scenario
of the structure formation.

\begin{figure}
\centerline{\includegraphics[width=0.8\textwidth]{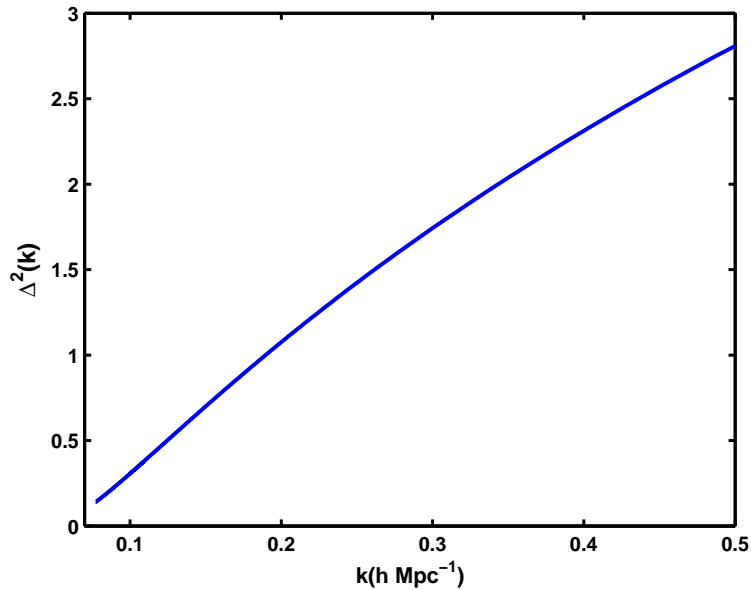}}
\caption{\small The power spectrum calculated by the two component
top-hat model is derived in terms of $k$.}
 \label{powerspectrum}
\end{figure}

An important observational data for examining this model is
comparing the power spectrum of the large scale structures at the
present time with that of model. We do this comparison by
calculating the mass variance for a sphere with the radius of $R$,
in terms of power spectrum. The total mass $M_R(r)$ in a sphere with
the radius $R$, centered on the point $r$ is:
\begin{eqnarray}
M_R(r) &=& \int^{|r-r'|<R}\rho(r')d^3r'\nonumber\\
&=&\int\rho(r')\Theta(R-|r-r'|)d^3r',
\end{eqnarray}
where $\Theta$ is the step function. Using the Fourier
transformation of density contrast and step function, the mass
variance is related to the power-spectrum through
\begin{equation}
\left(\frac{\Delta M_R}{M}\right)^2 = \int|W(kR)|^2 |\Delta_k|^2
\frac{dk}{k}, \label{massv}
\end{equation}
where $\Delta_k = {k^{3/2} |\delta_k|}/{\sqrt{2}\pi}$ and $W(kR)$
is the Fourier transformation of step function. The function
$W(kR)$ cuts off the integral (\ref{massv}) for $k>1/R$ and since
$\Delta_k$ is an increasing function of $k$, so the integral is
generally dominated at $k\sim1/R$ and we have \cite{peacock}:
\begin{equation}
\left(\frac{\Delta M_R}{M}\right)\sim\Delta_k ~~~~ k\sim 1.38/R.
\label{dm}
\end{equation}
Figure (\ref{powerspectrum}) shows the calculated power spectrum
in two component top-hat model in terms of $k$. This power
spectrum derived from this model has almost the same trend as the
observations \cite{peacock}. We should mention that since the
power-spectrum of galaxies has scale dependent biasing, these data
cannot be used for comparing with the dark matter spectrum. To
have a comparison of this model with the observed data, we use
$\sigma_8$ derived from the gravitational weak-lensing
observations \cite{ham03}.We point out that $\sigma_8$ derived
from the weak-lensing deosn't suffer from the biasing problem and
it probes directly the distribution of the dark matter. Weak
lensing  observations provide $0.62<\sigma_8<1.32$  which is
compatible with that of top-hat model, $\sigma_8\simeq 0.8$.

In the next section we will discuss about the effect of background
dynamics on the evolution of the structures namely the effect of a
variable dark energy model.

\section{Top-Hat Model in Variable Dark Energy Background}
\label{variable}
  In this section we discuss  the effect of background
dynamics on the growth rate of the structures. Recent observations
of SNIa and CMB show that universe is mainly made by an exotic
fluid so-called dark energy which speeds up its expansion
\cite{ris}. The standard solution for interpretation of the
positive acceleration of the universe is including the
cosmological constant to the Einstein equation. The best fit with
the observations provides $\Omega_m^{(0)} = 0.3$ and
$\Omega_{\Lambda}^{(0)}=0.7$, where $\Omega_{\Lambda}$ is the
density parameter corresponds to the cosmological constant. While
$\Lambda$CDM model provides a good fit to the SNIa and CMB data,
however it suffers the coincidence problem and finite tuning of
the cosmological constant at the early universe. One of the
solutions is considering a variable dark energy model. In this
model an evolving scalar field generates the energy and the
pressure of the dark energy and for the later times in the history
of universe, it provides a positive accelerating universe.

In this section we study the effect of variable dark energy on the
dynamics of the spherical collapse model as a function of the
redshift. The dark energy can influence on the growth of
large-scale cosmological structures through (i) the background
effect, in which the dark energy changes the expansion rate of the
background and (ii) dark energy may deviate from the homogenous
distribution due to the gravitational interaction with the dark
matter. The feedback of the dark energy is changing the growth
rate of the dark matter. These effects have been studied in the
following works
\cite{lahav91,wang98,iliv01,weinberg03,batt03,zeng04,mota04}.

In the case of $\Lambda$CDM universe the spherical collapse is
similar to that in CDM model, except the cosmological constant
that changes the growth of the structure though altering the
background dynamics. For the CDM model the virialization radius
$(R_{vir})$ is the half of the maximum radius of the structure
$(R_{max})$. In the $\Lambda$CDM universe the virialization radius
is smaller than that of in CDM model \cite{lahav91}. Mota and Van
de Bruck (MB) considered the spherical collapse for different
potentials of the quintessence models \cite{mota04}. They showed
that the predictions of the spherical collapse depend on the dark
energy model. In this scenario, during the collapse of over-dense
regions, the dark matter enters the highly non-linear regime while
the perturbation in dark energy deviates slightly from the
background. They also showed that if the dark energy equation of
state is assumed to be constant, the differences between the
homogenous and inhomogeneous cases are small. The advantage of
considering the dark energy in the spherical collapse model is
that it predicts that the cosmic structures such as the clusters
of galaxy have been collapsed prior to the epoch of $z \sim 1.4$,
compatible with the observations of the most distant cluster
\cite{basila06}.

\begin{figure}
\centerline{\includegraphics[width=0.8\textwidth]{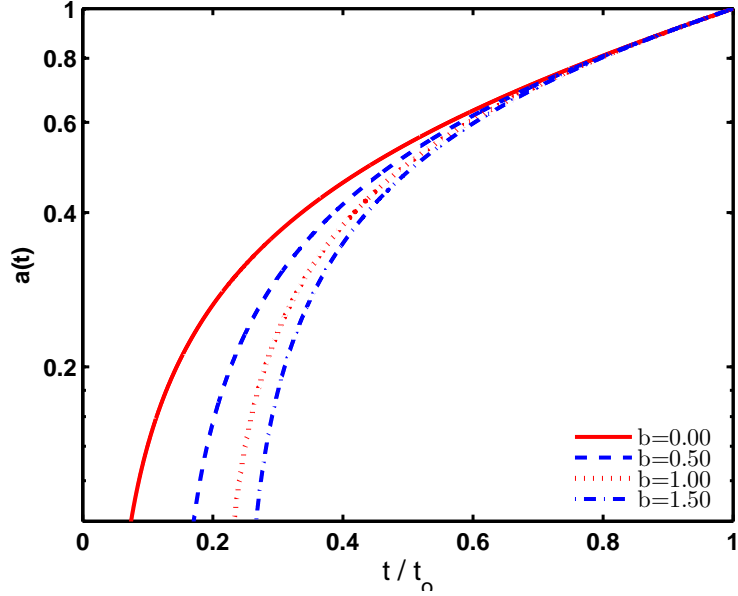}}
\caption{\small Dynamics of scale factor in terms of cosmic time
(normalized to $t_0$) with a variable dark energy model given by
Eq. (\ref{e_state}). Solid line stands for $b=0$ ($\Lambda$CDM),
dashed line $b=0.5$, dotted line $b=1.0$ and dashed-dotted line
$b=1.5$.}\label{adyn}
\end{figure}

\begin{figure}
\centerline{\includegraphics[width=0.8\textwidth]{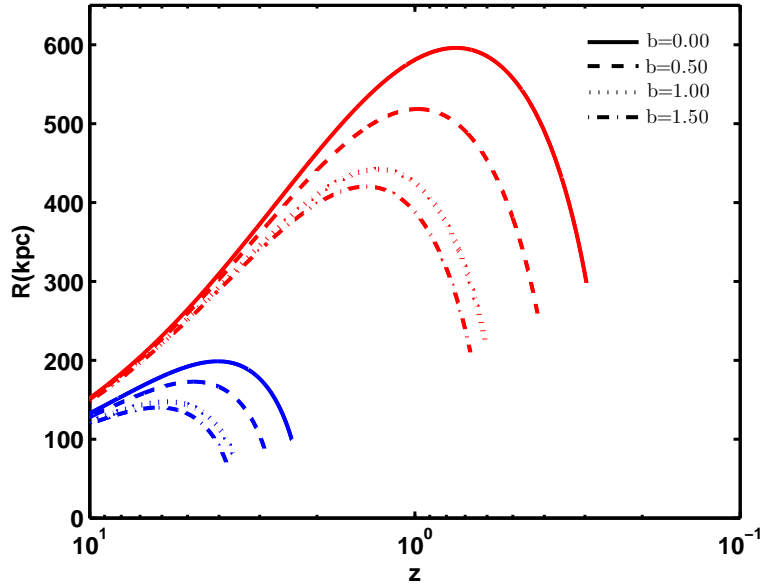}}
\caption{\small The effect of bending parameter in logarithmic
variable dark energy model on the dynamics of structure in terms
of redshift. Red lines represent the baryonic component and blue
lines stand for the dark matter component. The bending parameters
are chosen as $b=0$ (solid line), $b=0.5$ (dashed line), $b=1$
(dotted line) and $b=1.5$ (dashed-dotted line). The equation of
state is fixed with $\omega_{0}=-1.45$. } \label{rqb}
\end{figure}

\begin{figure}
\centerline{\includegraphics[width=0.8\textwidth]{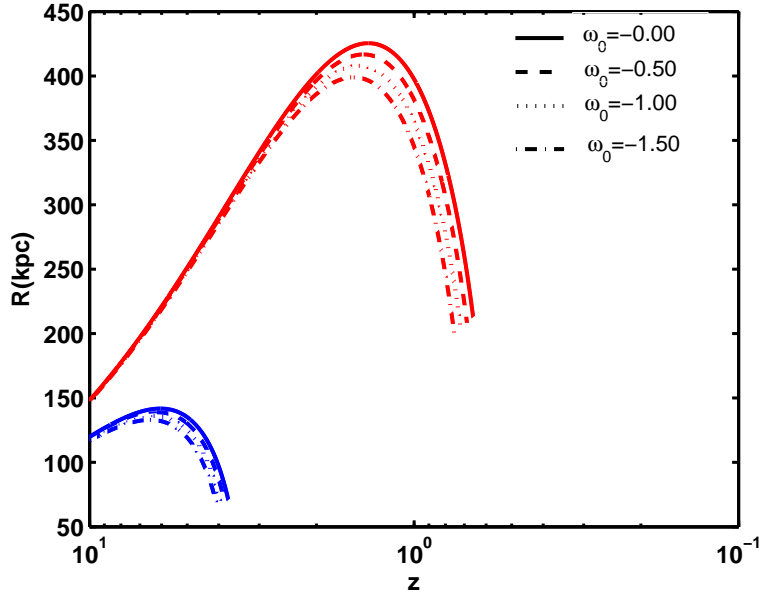}}
\caption{\small The effect of equation of state in logarithmic
variable dark energy model on the dynamics of structure in terms
of redshift. Red lines represent the baryonic component and blue
lines stand for the dark matter component. The equation of states
are chosen as $w_0=0$ (solid line), $w_0=-0.5$ (dashed line),
$w_0=-1$ (dotted line) and $w_0=-1.5$ (dashed-dotted line). The
bending parameter is fixed with $b=1.35$. } \label{rqw}
\end{figure}

In this part we examine the effect of a parameterized dark energy
on the evolution of the structures through the dynamics of the
background. We take the equation of state from Wetterich (2004):
\begin{equation}
{\omega}(z;b,\omega_{0})=\frac{\omega_{0}}{[1+b \ln(1+z)]^{2}},
\label{e_state}
\end{equation}
where $\omega_0$ is the equation of state at the present time and
$b$ is the bending parameter \cite{wet04}. The best fit from the
comparison of the cosmological data with the model results in $b =
1.35^{+1.65}_{-0.90}$ and $w_0 = -1.45^{+0.35}_{-0.60}$
\cite{mov06}. The density parameter of this dark energy from the
continuity equation changes as:
 \begin{equation}
 \Omega_{\lambda}(z;b,\omega_0)= \Omega_{\lambda}^{(0)}
 (1+z)^{3[1+\bar w(z;b,\omega_0)]},
 \end{equation}
where $\Omega_{\lambda}^{(0)}$ is the energy density of dark
energy at the present time and
${\bar\omega}(z;b,\omega_{0})=\omega_{0}/[1+b \ln(1+z)]$ is the
average of the equation of state in the logarithmic scale. Using
the Hubble parameter for the flat universe,
\begin{equation}\label{hub7}
H^2(z;b,\omega_0)=H_0^2[\Omega_m^{(0)}(1+z)^3+\Omega_{r}^{(0)}(1+z)^4+\Omega_{\lambda}^{(0)}(1+z)^{3(1+\bar
w)}],
\end{equation}
we obtain the dynamics of the scale factor for the various bending
parameters as shown in Figure (\ref{adyn}). To calculate the
evolution of the baryonic and dark matter components of the
structure, we use equations (\ref{dark}) and (\ref{baryon}) to
obtain the radius of the structure as a function of time. The
evolution of the Hubble parameter as a function of redshift is given
by equation (\ref{hub7}). Eliminating time in favor of the redshift
for the dynamics of the structures we obtain
$R = R(z)$ as shown in Fig. (\ref{rqb}). Here we fix the value of
$w_0$ and calculate the dynamics of the structures for various
bending parameters. Increasing the bending parameter causes the
structure forms at earlier epochs but with smaller radius. Also
the effect of $w_0$ on the formation of structures, while $b$ is a
fixed value, is shown in Fig. (\ref{rqw}). Decreasing $\omega_{0}$
causes the structures form faster with smaller radius.

In the dark energy models, the evolution of the structures deviate
from that in $\Lambda$CDM model. This effect results from the
change in the background dynamics due to the dark energy.
In addition to this effect, the total energy of the structure will
be changed due to the dark energy effect. The effect of a variable
dark energy in contrast to $\Lambda$CDM model is that we can have a
none-zero contribution of the dark energy at the early epoches of
the universe. So for a given $H_0$ at the present time, we expect to
have larger $H_i$ at the early times. Looking to the total energy of
a structure $E = -3/2H_i^2R_i^2\delta_i$ shows that having larger
$H_i$ causes more bounded structure. This effect provides a negative
energy for the structure which will produce smaller $R_{max}$ and
$R_{vir}$ as well as earlier virialization time to the structure.

In the next section we will study the effect of cooling of baryonic
matter at the finial stage of virialization on structure formation.

\section{Cooling Mechanism}
\label{cooling} Once the baryonic structure reaches to the virial
condition, radial velocity of the structure converts to the
dispersion velocity or in anther word the baryonic gas
thermalizes. The temperature of structure, using the kinetic
energy of baryonic particles can rise up to $10^7$K and from the
Saha equation we will have an ionized medium. The ionized plasma
then cools down and the result is more contraction of the baryonic
structure. On the other hand through the cooling, the baryonic
structure can fragment into the small parts to generate stellar
systems. During the cooling of baryonic matter, it will lose its
kinetic energy and hence falls into the gravitational potential
well and subsequently gains the kinetic energy \cite{nul95}. This
cooling and heating continues until the system reaches to a
quasi-steady state.

A simple parameter that represents the cooling of a gas is the
ratio of cooling to the free fall time scale,
$\tau={t_{cool}}/{t_{grav}}$. The cooling time is defined by
$t_{cool} = E/{\dot E}\approx{3\rho k_{B}T}/{2\mu\Lambda(T)}$ and
the dynamical time also results from the time scale for the free
fall of a structure and is given by $t_{dyn} =
\frac{\pi}{2}(2GMR^{-3})^{-1/2}$. If $\tau>1 $ then the gas can
cool; but as it cools the gas can retain the pressure support by
adjusting its pressure distribution. If $\tau<1$ the gas will cool
rapidly to a minimum temperature. This will lead to the loss of
pressure support and the gas will undergo an almost free-fall
collapse. In this case fragmentation into stellar units can occur.
There are various physical processes contribute in cooling,
depending on the temperature of the plasma. We assume cooling is
dominated by plasma Bremsstrahlung radiation and recombination. At
temperature larger than $10^7 K$, Bremsstrahlung dominates the
cooling whereas in the range of $10^4\sim10^6$, recombination of
gas is the main source of cooling. The net cooling rate is
\begin{equation}
C=\frac{dE}{dtdV}=n_{e}n_{p}\Lambda(T),
\end{equation}
 where $\Lambda(T)$ is the radiative cooling function and is expressed as
\begin{equation}
\Lambda(T)=(A_{B}T^\frac{1}{2}+A_{R}T^\frac{-1}{2})\rho^2 ~~~
\frac{erg}{cm^3s}.
\end{equation}
$A_{B}\propto e^6/m_e^{3/2}$ represents cooling due to the
bremsstrahlung and $A_{R}\simeq e^4m_pA_B$ arises from the
recombination. This expression is valid for temperatures above
$10^4~ K$. For lower temperatures, the cooling rate drops
drastically because H can no longer be significantly ionized by
collisions. The radiative cooling function for temperatures above
$10^{7}$ is well approximated by
\begin{equation}
\Lambda(T)=2.5\times10^{-27}T^\frac{1}{2}\frac{erg}{cm^3s}.
\end{equation}

\begin{figure}
\centerline{\includegraphics[width=0.8\textwidth]{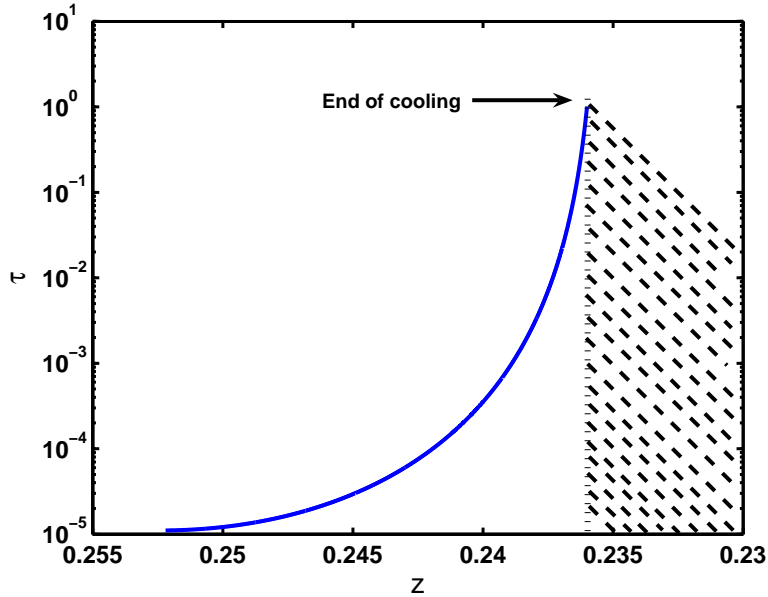}}
\caption{\small Variation of $\tau$ in term of redshift for a
galaxy mass structure. $\tau$ is plotted from thermalization till
the quasi-steady state phase of baryonic structure. Dashed era
represents the stable zone for the structure where the cooling is
negligible.} \label{cool1}
\end{figure}

\begin{figure}
\centerline{\includegraphics[width=0.8\textwidth]{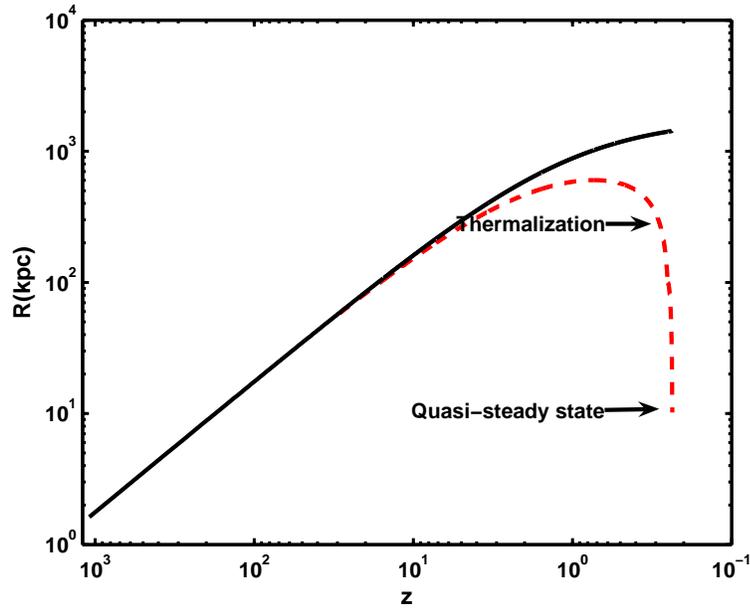}}
\caption{\small Dynamics of radius of galaxy mass structure in
terms of redshift (dashed line). After thermalization we will have
free fall collapse of structure until quasi steady state time.
Solid line represents the dynamics of background for comparison.}
\label{cool2}
\end{figure}

\begin{figure}
\centerline{\includegraphics[width=0.8\textwidth]{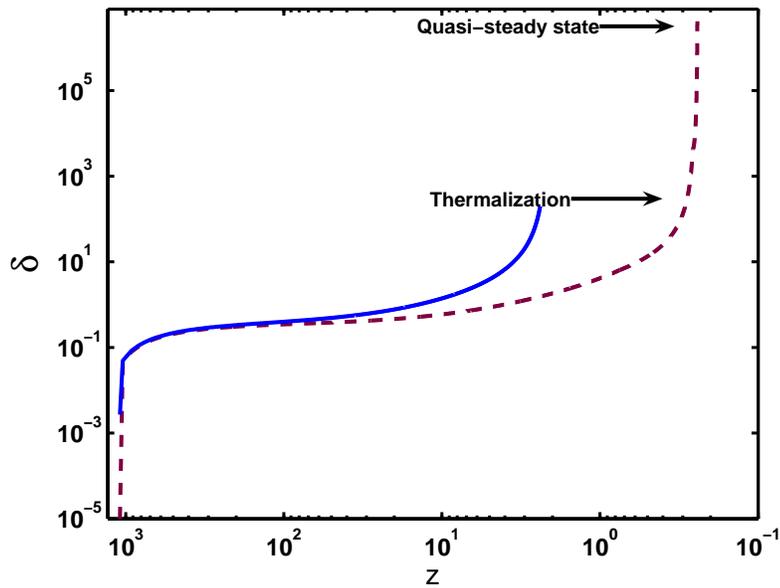}}
\caption{\small Evolution of density contrast in terms of redshift
for baryonic (dashed line) and dark matter (solid line) components
of a galaxy mass structure. After thermalization we will have
almost four order of magnitude increase of the baryonic density
contrast.} \label{cool3}
\end{figure}

Here we study the cooling condition of baryonic part of structure
after the thermalization which $\tau=t_{cool}/t_{dyn}\ll1$ is hold
and we have almost free fall condition for the structure. Fig.
(\ref{cool1}) shows the variation of $\tau$ in terms of redshift
for a galaxy mass structure, starting from the thermalization time
to the quasi steady state. The duration of cooling for this
structure in the $\Lambda$CDM model is about $\Delta z \simeq
0.07$. Fig. (\ref{cool2}) shows the evolution of the size of
structure in terms of redhsift for this period. After
thermalization structure freely falls until it reaches to a quasi
steady state. Fig. (\ref{cool3}) also indicates the evolution of
the density contrast including the cooling effect after
thermalization time.

To see the effect of variable dark energy on cooling time, we take
a small structure with the mass of $10^6 M_{\odot}$. This mass is
suitable for studying the evolution of the globular clusters and
corresponds to the baryonic Jeans mass after the decoupling
\cite{peeb68}. According to the structure formation scenario we
expect that this structure has a dark matter component. However
the observations of dispersion velocity show that these structures
almost have no dark matter. The pre-galactic model for the
globular cluster and tidal striping of dark halo by Galaxy can
explain the lack of dark matter in these structures \cite{ros88}.
It should be noted that in the formation of these structures we
don't consider merging effect as naturally is taken account in the
N-body simulations.

Variable dark energy causes the structure thermalizes and cools at
the higher redshifts (see table \ref{tab2}). For instance, taking
the parameters $w_0=-1.45$ and $b=1.35$, structures with globular
cluster mass thermalize at $z\sim 2.17$ and stop cooling at $z\sim
1.94$. For a larger structure with a galaxy mass of $10^{11}
M_{\odot}$, the corresponding thermalization redshift occurs at
$z\sim 0.7$ (see Fig. \ref{rqb}). Comparing with the globular
cluster we can conclude that the first star bursts would happened
in the globular clusters. The high metalicity with the low
rotation of globular cluster in the galactic halo supports this
hypothesis.

\section{Conclusion}
\label{conc} In this work we extend simple top-hat model into two
component dark matter-baryonic structure. The initial condition
for the sub-horizon size structures is taken at the last
scattering surface. The scale of $\lambda<d_H$ for the structures
guarantees applying approximately the Newtonian mechanics. Using
the Harrison-Zeldovich spectrum for the perturbations of dark
matter, we obtained the evolution of each component of the
structure.

For the dark matter part, we showed that density contrast in the
small mass structures grows faster than the larger ones and
subsequently reaches to the maximum radius and virializes at the
higher redshifts. This behavior of structures implies that the
star burst should take place at the smaller structures as dwarf
galaxy and globular clusters. An observable parameter of
structures to compare with the model is the power spectrum. We
calculated the mass variance of structures in various scales at
the present time and compared $\sigma_ 8\sim 0.8$ from the model
with that of observation from the weak lensing. For studying the
dynamical effect of background on the evolution of the structures,
we applied a logarithmic variable dark energy model and showed
that the structures in this model evolve faster than that of
$\Lambda$CDM.

\newpage

\newpage
\newpage
\newpage
\newpage

\newpage

\newpage
\newpage
\newpage

\newpage
\newpage

\newpage

\begin{table}
\begin{center}
\caption{Numerical results from the evolution of structures with
various masses. First column shows the range of mass of structures
and subscript in $M$ represents mass in $10^{n} M_{\odot}$ .
Second column is the corresponding redshift to entering to the
Horizon. The density contrast of dark matter at the decoupling
epoch is in third column. Forth and Fifth columns are the
corresponding redshift of maximum radius for the dark matter and
virialization of the dark structure. Sixth and seventh columns are
the same as the forth and fifth columns for the baryonic
structure.} {\begin{tabular}{|c|c|c|c|c|c|c|}
\hline
M & $z_{enter}$ & $\delta_{dec}^{d}$ & $z_{m}^{d}$ & $z_{vir}^{d}$ & $z_{m}^{b}$ & $z_{vir}^{b}$\\
\hline
$M_{15} $ &$5.90\times10^{3}$&$1.9\times10^{-4}$&0.70&0.30&0.15&0.10\\
$M_{14}$&$1.27\times10^{4}$&$1.2\times10^{-3}$&1.70&0.70&0.26&0.15\\
$M_{13}$ &$2.74\times10^{4}$&$2.0\times10^{-3}$&2.60&1.50&0.40&0.20\\
$M_{12}$ &$5.90\times10^{4}$&$2.7\times10^{-3}$&4.00&2.30&0.75&0.30\\
$M_{11}$ &$1.27\times10^{5}$&$3.4\times10^{-3}$&5.20&3.20&1.13&0.50\\
$M_{10}$ &$2.74\times10^{5}$&$4.2\times10^{-3}$&6.70&4.20&1.60&0.78\\
$M_{9}$ &$5.90\times10^{5}$&$4.9\times10^{-3}$&8.00&5.00&2.00&1.00\\
$M_{8}$ &$1.27\times10^{6}$&$5.6\times10^{-3}$&9.30&5.90&2.40&1.20\\
$M_{7}$ &$2.74\times10^{6}$&$6.3\times10^{-3}$&10.50&6.70&2.80&1.40\\
$M_{6}$
&$5.90\times10^{6}$&$7.1\times10^{-3}$&12.00&7.75&3.30&1.80\\
\hline
\end{tabular}
\label{tab1}}
\end{center}
\end{table}

\newpage
\begin{table*}
\begin{center}
\caption{The effect of bending parameter on duration of cooling of
baryonic structure in logarithmic variable dark energy model. Mass
of structure is taken $10^6 M_{\odot}$. At the first column we
fixed $(w_0=-1.45)$ and the second column shows the duration of
cooling for various bending parameters. In the third column we
fixed bending parameter $(b=1.35)$ while the equation of state
$w_0$ has different values with  corresponding duration of
redshift is indicated in the fourth column.}
\begin{tabular}{|c|c|c|c|}
\hline
$\omega_{0}=-1.45$ & $z_{star burst}$& $b=1.35$ & $z_{star burst}$  \\
\hline
$b=0.00 $ &$[1.82,1.65]$&$\omega_{0}=0.00$&$[1.80,1.62]$\\
$b=0.50$&$[2.15,1.84]$&$\omega_{0}=-0.50$&$[2.04,1.80]$ \\
$b=1.00$ &$[2.22,1.98]$&$\omega_{0}=-1.00$ &$[2.12,1.90]$\\
$b=1.50$ &$[2.25,2.00]$&$\omega_{0}=-1.50$ &$[2.19,1.95]$\\
\hline
\end{tabular}
\label{tab2}
\end{center}
\end{table*}


\begin{thebibliography}{0}


\bibitem{smo92}
G. F. Smoot {\it et al.}, ApJ {\bf 396}, 1 (1992)

\bibitem{hin06}
G. Hinshaw {\it et al.}, ApJS {\bf 170}, 263 (2007).


\bibitem{spe03}
D. N. Spergel  {\it et al.}, ApJS {\bf 148}, 175 (2003).


\bibitem{spe06}
D. N. Spergel {\it et al.}, ApJS {\bf 170}, 377 (2007).


\bibitem{spri05}
V. Springel {\it et al.}, 2005, Nature, 435, 629.

\bibitem{pee80}
P. J. E. Peebles 1980, The Large-Scale Structure of the Universe
(Princeton University Press, Princeton, New Jersey).


\bibitem{sil68}
J. Silk, ApJ {\bf 151}, 459 (1968)

\bibitem{man02}
R. Mansouri and S. Rahvar, IJMPD {\bf 11}, 321 (2002).




\bibitem{rah03}
S. Rahvar, IJMPD {\bf 12}, 79 (2003).


\bibitem{pad93}
T. Padmanabhan, 1993, Structure Formation in the Universe,Cambridge
Univ. Press.


\bibitem{pad00}
T. Padmanabhan, 2002, Theoretical Astrophysics, Vol III. Cambridge
Univ. Press.



\bibitem{har70}
E. R. Harrison, Phys. Rev. D {\bf 1}, 2726 (1970)

\bibitem{zeld70}
Y. Zeldovich, Astr. Astron. {\bf 5}, 8 (1970)

\bibitem{peacock}
J. A. Peacock, 1999, Cosmological Physics. Cambridge Univ. Press


\bibitem{ham03}
T. Hamana, ApJ{\bf 597}, 98 (2003)

\bibitem{ris}
A. G. Riess  {\it et al.}, ApJ {\bf 607}, 665 (2004)

\bibitem{lahav91}
O. Lahav, P. B. Lilje, J. R. Primack, M. J. Rees, MNRAS. {\bf
251}, 128 (1991).

\bibitem{wang98}
L. M. Wang, P. J. Steinhardt, ApJ. {\bf 508}, 483 (1998).

\bibitem{iliv01}
I. T. Iliev, P. R. Shapiro, MNRAS. {\bf 325}, 468 (2001).

\bibitem{weinberg03}
N. N. Weinberg, M. Kamionkowski, MNRAS. {\bf 341}, 251 (2003).

\bibitem{batt03}
R. A. Battye, J. Weller, Phys. Rev. D {\bf 68}, 083506 (2003).
\bibitem{zeng04}
D. F. Zeng, Y. H. Gao, arXiv:astro-ph/0412628.

\bibitem{mota04}
D. F. Mota, C. van de Bruck, Astron. Astrophys. {\bf 421}, 71
(2004).


\bibitem{wet04}
C. Wetterich, Phys. Lett. B {\bf 594},17 (2004)



\bibitem{mov06}
M. S. Movahed, S. Rahvar, Phys. Rev. D {\bf 73}, 083518 (2006)



\bibitem{nul95}
P. E. J. Nulsen, A. C. Fabian, MNRAS {\bf 271}, 561 (1995)



\bibitem{peeb68}
P. J. E. Peebles and Dicke, R. H., ApJ {\bf 154}, 891 (1968)


\bibitem{ros88}
E. I. Rosenblatt., S. M. Faber, and G. R. Blumenthal, ApJ {\bf 330},
191 (1998).



\bibitem{basila06}
S. Basilakos, N. Voglis, MNRAS, {\bf 374}, 269 (2007).
\end{thebibliography}
\end{document}